\documentclass[12pt]{article}
\begin{document}
\title{On the Photon Mass}
\author{B.G. Sidharth$^*$\\
Frankfurt Institute of Advanced Studies\\
Max Von Laue Strasse 1, D-60438, Frankfurt am Main, Germany\\
$^*$ Permanent Address: International Institute for Applicable Mathematics\\
\& Information Sciences\\
Hyderabad (India) \& Udine (Italy)\\
B.M. Birla Science Centre, Adarsh Nagar, Hyderabad - 500 063
(India)}
\date{}
\maketitle
\begin{abstract}
We review the case for the photon having a tiny mass compatible with
the experimental limits. We go over some possible experimental tests
for such a photon mass including the violation of Lorentz symmetry.
We point out that such violations may already have been witnessed in
tests involving high energy gamma rays from outer space as also
ultra high energy cosmic rays.
\end{abstract}
\section{Introduction}
Generally the mass of the photon is taken to be zero. What is not so
well known is the fact that a photon with a non zero mass does not
contradict established theory, though experiments have provided
upper limits to such a mass. These upper limits have become more and
more stringent and presently a good estimate seems to be
$10^{-57}gms$. The author's work has shown that, on theoretical
grounds, the photon should have a mass $\sim 10^{-65}gms$
\cite{bhtd}, which is below the experimental limit. We will come
back to the theoretical framework which suggests the above photon
mass, but before that we will discuss some experimental tests which
can confirm the above results. Indeed it must be mentioned that
already experiments suggest a non vanishing photon mass.
\section{Experiments and Results}
1. Laboratory tests have suggested that Maxwell's displacement
current is a real phenomenon \cite{vigier,bartlett}, which leads to
the fact that vacuum contains a dissipative mechanism or friction,
which again leads to a non zero photon mass. This was achieved by
Bartlete and Corle by using squids which enable them to directly
measure the magnetic field inside a thin circular parallel plate
capacitor, as it was being charged.\\
2. We now briefly describe an experiment which so far has yielded
limits \cite{lakes}, in agreement with other experimental limits. In
this experiment, described by Lakes an influence of large cosmic
magnetic vector fields on massive photons can be measured in the
laboratory via the energy density of $\mu_\gamma A^2$, where
$\mu_\gamma$ is the Compton wavelength of the photon. A modified
Cavendish balance was used for this purpose. A toroid of electrical
steel was wound with a current carrying wire and supported by water
floatation. Stability and a restoring torque was provided by a
tungsten wire. A fine copper wire provided an electrical return
path. Further a magnetic shield of MU metal was added to try and
eliminate noise due to magnetic field fluctuations. The operative
relation is given by
\begin{equation}
\mu^2_\gamma |A_{\mbox{ambient}}| = \frac{G\frac{1}{L} \frac{\pi
d^4}{32}\frac{\Phi}{2}}{k\left[\frac{1}{4}(w - u)^2 nhI
ln\left(\frac{w}{w-u}\right)\right]}\times sin (\Theta_A)\label{e1}
\end{equation}
in which $\Theta_A$ is on the angle between $A_{\mbox{ambient}}$ and
Earth's rotation axis. As can be seen from (\ref{e1}) the torque
vanishes if the photon does not have a mass. Alternatively this
could happen if the cosmic ambience vector potential were
accidentally aligned with the earth's axis of rotation. Such an
accident was of course ruled out. The experiment with values
substituted into (\ref{e1}) suggested that the photon mass is less
than
$10^{-52}gms$.\\
3. Mignani and co-workers have performed experiments which
demonstrate, surprisingly a violation of Lorentz Invariance in the
low energy regime \cite{cms}. In fact this is exactly what we would
expect with the photon mass suggested above. This experiment was
first performed in 1998 and then repeated in 1999 with improved
sensitivity. Both gave the same result.\\
The experiment attempted to detect a DC voltage across a conductor
induced by the steady magnetic field of a coil. Such a voltage would
be a signature of a violation of Lorentz Invariance, as it would
indicate a non zero Lorentz force between the magnetic field
generated by a stationary current and a charge, both of which are at
rest in the same inertial reference frame. If Lorentz Invariance was
strictly valid, then such a force would not have existed. The
conclusion was that there was such a Lorentz violation at an energy
threshold of $10^{-9}eV$, corresponding to a low energy regime. It
may be pointed out that this threshold is in fact above that in a
massive photon with the mass suggested above.\\
4. With a non zero photon mass we would have, for radiation
\begin{equation}
E = h \nu = m_\nu c^2 [1 - v^2_\gamma /c^2]^{-1/2}\label{ea1}
\end{equation}
From (\ref{ea1}) one would have a dispersive group velocity for
waves of frequency $\nu$ given by (Cf. also ref.\cite{vigier})
\begin{equation}
v_\gamma = c \left[1 - \frac{m^2_\gamma c^4}{h^2
\nu^2}\right]^{1/2}\label{ea2}
\end{equation}
We would like to point out that (\ref{ea2}) indicates that higher
frequency radiation has a velocity greater than lower frequency
radiation reaching one thousandth of a second earlier. This is a
very subtle and minute effect and is best tested in for example, the
observation of high energy gamma rays, which we receive from deep
outer space. It is quite remarkable that we may already have
witnessed this effect- higher frequency components of gamma rays in
cosmic rays do indeed seem to reach earlier than their lower
frequency counterparts \cite{pav}. The GLAST satellite of NASA to be
launched later in 2007 may be able to
throw more light on these high energy Gamma rays.\\
5. Another consequence of the non zero photon mass is that the usual
Coulomb potential goes over into a Yukawa potential with a range
equal to the Compton length of the photon which in fact is $\sim
10^{28}cms$ which again is the radius of the universe. So the effect
will be very small indeed but in principle this would lead to, for
example a shift in the fine structure energy levels that could be
observed.\\
Let us introduce the Yukawa potential,
\begin{equation}
V(r) = \alpha e^{-\mu r}/r\label{eb1}
\end{equation}
into the Dirac equation instead of the usual Coulomb potential
\cite{greiner2}. In (\ref{eb1}), $\mu$ is proportional to the mass
of the photon $\sim 10^{-65}gms$ and is therefore a very small
quantity. We introduce (\ref{eb1}) into the stationary Dirac
equation to get
\begin{equation}
[c \vec{\alpha} \cdot \vec{p} + \beta m_0c^2 - (E - V(r))] \psi (r)
= 0\label{eb2}
\end{equation}
From (\ref{eb2}) and (\ref{eb1}), we can immediately see that roughly the effect of the photon mass is to shift the energy levels by a miniscule amount.\\
We further introduce the notation
\begin{equation}
Q = 2 \lambda r, \quad \mbox{where} \, \lambda = \frac{\sqrt{m^2_0
c^4 - E^2}}{hc},\label{eb3}
\end{equation}
After the standard substitutions (Cf.ref.\cite{greiner2}) we finally
obtain
$$\frac{d\Phi_1}{dQ} = \left(1 - \frac{\alpha E}{hc\lambda Q}\right) \Phi_1 - \left(\frac{\chi}{Q} + \frac{\alpha m_0c^2}{hc\lambda Q}\right) \Phi_2,$$
\begin{equation}
\frac{d\Phi_2}{dQ} = \left(- \frac{\chi}{Q} + \frac{\alpha
m_0c^2}{hc\lambda Q}\right) \Phi_1 + \left(\frac{\alpha E}{hc\lambda
Q}\right) \Phi_2\label{eb4}
\end{equation}
The substitutions
$$\Phi_1 (Q) = Q^\gamma \sum^{\infty}_{m=0} \alpha_m Q^m,$$
\begin{equation}
\Phi_2 (Q) = Q^\gamma \sum^{\infty}_{m=0} \beta_m Q^m.\label{eb5}
\end{equation}
in (\ref{eb4}) leads to
$$\alpha_m (m+\gamma) = \alpha_{m-1} - \left(\frac{\alpha E}{hc\lambda}\right) \alpha_m -
\left(\chi + \frac{\alpha m_0c^2}{hc\lambda}\right)\beta_m,$$
\begin{equation}
\beta_m (m+\gamma) = \left(-\chi + \frac{\alpha
m_0c^2}{hc\lambda}\right) \alpha_m + \left(\frac{\alpha
E}{hc\lambda}\right)\beta_m.\label{eb6}
\end{equation}
As is well known $\gamma$ in (\ref{eb5}) is given by
\begin{equation}
\gamma = \pm \sqrt{\chi^2 - \alpha^2},\label{eb7}
\end{equation}
where $\lambda$ is given by (\ref{eb3}).\\
At this stage we remark that the usual method adopted for the
Coulomb potential is no longer valid - mathematically, Sommerfeld's
polynomial method becomes very complicated and even does not work
for a general potential: We have to depart from the usual procedure
for the Coulomb potential in view of the Yukawa potential
(\ref{eb1}). Nevertheless, it is possible to get an idea of the
solution by a slight modification. This time we have from
(\ref{eb6}), instead the equations
$$\alpha_m (m+\gamma) = \alpha_{m-1} - \frac{\alpha E}{\hbar c\lambda} \alpha_m +
\frac{\alpha E \mu}{\hbar c\lambda} \alpha_{m-1}$$
\begin{equation}
(-\chi \beta_m) - \frac{\alpha m_oc^2}{\hbar c\lambda} \beta_m +
\frac{\mu \alpha m_0c^2}{\hbar c\lambda} \beta_{m-1}\label{eb8}
\end{equation}
and
$$(m + \gamma) \beta_m = - \left(\chi + \frac{\alpha m_0c^2}{\hbar c \lambda}\right)
\alpha_m - \frac{\mu \alpha m_0c^2}{\hbar c\lambda} \alpha_{m-1}$$
\begin{equation}
+ \frac{\alpha E}{\hbar c \lambda} \beta_m - \frac{\mu \alpha
E}{\hbar c \lambda}  \beta_{m-1}\label{eb9}
\end{equation}
After some algebra on (\ref{eb8}) and (\ref{eb9}) we obtain
\begin{equation}
P \alpha_m + Q \beta_m = R \alpha_{m-1}\label{eb10}
\end{equation}
\begin{equation}
S \alpha_m + T \beta_m = U \beta_{m-1}\label{eb11}
\end{equation}
where $P,Q,S,T$ can be easily characterized, in the derivation of which we will neglect
$\mu^2$ and higher orders.\\
If
$$\alpha_m / \beta_m \equiv p_m$$
then we have from (\ref{eb10}) and (\ref{eb11}),
$$S p_m + T = \frac{U (QS-PT)}{(RSp_{m-1} - PU)}$$
We note that the asymptotic form of the series in (\ref{eb5}) will
not differ much from the Coulomb case and so we need to truncate
these series. For the truncation of the series we require
$$QS = PT$$
This gives
$$\left\{1 + \frac{\chi \hbar c \lambda}{\alpha m_0c^2}\right\} \left[E\left(\chi +
\frac{\alpha m_0c^2}{\hbar c \lambda}\right) + m_0c^2 \left(m+\gamma - \frac{\alpha E}
{\hbar c \lambda}\right)\right]$$
$$ = \left(m+\gamma - \frac{\alpha E}{\hbar c \lambda}\right) \frac{\hbar c \lambda}
{\alpha m_0c^2} \left[E\left(m+\gamma + \frac{\alpha E}{\hbar c \lambda}\right) + m_0c^2
\left(\chi + \frac{\alpha m_0c^2}{\hbar c\lambda}\right)\right]$$
Further simplification yields
$$(m+\gamma)^2 + \left\{\frac{\chi \hbar c \lambda + \alpha m_0c^2}{\hbar c \lambda}
\right\}^2 + \frac{\alpha^2 E^2}{\hbar^2c^2\lambda^2}$$ where
$\gamma$ is given by (\ref{eb7}). Finally we get, in this
approximation,
\begin{equation}
E^2 = m_0^2 c^4 \left[1 - \frac{2\alpha^2}{\alpha^2 + (m+\gamma)^2 -
\chi^2}\right] + A O(\frac{1}{m^2})\label{eb12}
\end{equation}
In (\ref{eb12}) $A$ is a small quantity
$$A \sim m^2_0 c^4 - E^2$$
The second term in (\ref{eb12}) is a small shift from the usual
Coulombic energy levels. In (\ref{eb12}) $m$ is a positive integer
and this immediately provides a comparison with known fine structure
energy levels. To see this further let us consider large values of
$m$. (\ref{eb12}) then becomes
\begin{equation}
E = m_0c^2 \left[1 - \frac{\alpha^2}{m^2_1}\right]\label{eb13}
\end{equation}
while the usual levels are given by
\begin{equation}
E = m_0c^2 \left[ 1 - \frac{\alpha^2}{2m^2_2}\right]\label{eb14}
\end{equation}
We can see from (\ref{eb13}) and (\ref{eb14}) that the photon mass
reproduces all the energy levels of the Coulomb potential but
interestingly (\ref{eb13}) shows that there are new energy levels
because $m_1^2$ in the new formula can be odd or even but $2m^2_2$
in the old formula is even. However all the old energy levels are
reproduced whenever $m^2_1 = 2m^2_2$.
If $\mu = 0$, then as can be seen from (\ref{eb2}), we get back the Coulomb problem.\\
Finally it may be remarked that as noted, in general it is not
possible to use  Sommerfeld's polynomial method for the Dirac
equation and the Coulomb potential is a special case \cite{bgs}.
Similarly, in scattering problems, the consequence of the Yukawa
potential would be that we can use the usual phase shift analysis
for short range (i.e. faster than Coulomb) potentials as the time
$\delta_e/\delta_{l-1}$ would converge, though slowly, with
increasing $l$.\\
6. Interestingly there is a cosmic coincidence which has puzzled
astronomers and which can be explained exactly by the photon mass
\cite{mp}. There has been a wealth of data from the WMAP (Cf. for
example \cite{ben,hin}). One of the intriguing findings is that the
dark energy domination and the CMB power supression, both occur
around the same red shift and energy scale - corresponding to the
Hubble radius $\sim 10^{-33}eV$. As has been pointed out \cite{mer},
this raises three disturbing questions, viz., the small magnitude,
the so called tuning problem and why this should occur during this
epoch, that is the coincidence problem, and finally why do both
coincidences occur at the energy scale of our present Hubble radius
of $10^{-33}eV$. The question has also been asked, "Does all this
suggest new physics?".\\
We now show that this is explained in terms of the Planck scale
underpinning for the universe, as has been discussed in detail in
the literature (Cf. refs.\cite{fpl1,fpl2,ijmpe,uof} and references
therein) and indeed is an observational confirmation for the model.
We summarize the main results. The universe can be considered to
have an underpinning of $N$ Planck oscillators. Further we have
\begin{equation}
R = \sqrt{N} l_P\label{ec1}
\end{equation}
In (\ref{ec1}) $R (\sim 10^{-33}eV)$  is the radius of the universe
and $l_P \sim 10^{-33}cms$ is the Planck length. It can also be
shown that $N \sim 10^{120}$. Moreover there is a minimum mass (or
minimum dark energy scale) in the universe which is given by
\begin{equation}
m = m_P / \sqrt{N} \sim 10^{-65}gm \sim 10^{-33}eV\label{ec2}
\end{equation}
(Details can be found in the above references).\\
It can now be seen that the puzzling small energy scale $\sim 10^{-33}eV$ referred to
earlier is exactly the same as the minimum allowable mass $m$ given in (\ref{ec2}).
Moreover the Compton wavelength of this mass is exactly $\sim R$, the Hubble radius
referred to earlier. What all this means is, at this point of time in the universe,
there is a minimum energy of the background dark energy $\sim 10^{-33}eV$.\\
7. Finally we remark on some indirect evidence for the photon mass.
It has been shown by Ignatiev and Joshi \cite{ig} that a photon with
a vanishing mass would imply the existence of the magnetic monopole.
The fact that no monopole has been found even after nearly eighty
years should therefore indicate that the photon has a small mass.
\section{Theoretical Framework}
We now come to the theoretical framework which prompted the author
to suggest the above photon mass \cite{bhtd,mp}. Recently the author
deduced from theory that the photon has a small mass of $\sim
10^{-65}gms$, this being consistent with the velocity of light and
also with experimental limits on the photon mass as above discussed
already \cite{bhtd}. Interestingly from purely thermodynamic
reasoning Landsberg had shown that the above mass is the minimum
allowable mass in the universe \cite{land}.  More recently from a
completely unrelated point of view, it was shown by the author that
this is the minimum allowable mass in the universe in a model where
zero point
oscillators at the Planck scale provide the underpinning for the universe \cite{found,uof}.\\
The derivation of the photon mass uses the fact that for the
Langevin equation in the limiting case of low viscous resistance, a
particle behaves like a Newtonian particle moving in the absence of
any external forces with a uniform velocity given by
$$\langle v^2 \rangle = \frac{kT}{m}$$
If in the above relation we use extreme values of the minimum
thermodynamic temperature of the universe and the minimum mass, we
recover the velocity of light. To be more specific we use the
Beckenstein temperature given by
$$T = \frac{\hbar c^3}{8\pi kMG}$$
with $M \sim 10^{55}gms$, the known mass of the universe. This gives
us the value
$$T \sim 10^{-28}K$$
For the minimum mass we use a result due to Landsberg
(Cf.ref.\cite{land}), and also the same result independently
obtainable as noted above from a model of Planck oscillators, viz.,
$$m \sim 10^{-65}gms$$
It must be stressed that though the above equation for the average
velocity square resembles the root mean square temperature equation
of thermodynamics, it is in fact different. Moreover, this velocity
would be maintained for the age of the universe and would be Lorentz
invariant \cite{evans} (Cf.ref.\cite{bhtd} for details). It may be
pointed out that all this is in the spirit of the experimental
results of Bartlete and Corle referred to in Section 2.\\
We now give a further theoretical justification for the above. We
first observe that as is well known \cite{neeman}, Maxwell's
equations can be written in the following form
\begin{equation}
{\bf \Psi} = \vec{E} + \imath \vec{B},\label{ed3}
\end{equation}
\begin{equation}
\vec{\nabla} \times {\bf \Psi} = \imath \dot{{\bf \Psi}}\label{ed4}
\end{equation}
\begin{equation}
\vec{\nabla} \cdot {\bf \Psi} = 0\label{ed5}
\end{equation}
Equations (\ref{ed3}) to (\ref{ed5}) will be useful in the sequel.\\
We next observe that Maxwell's equations have been deduced in a
fashion very similar to the Dirac equation, from first principles
\cite{gersten}. In this deduction, we use the usual energy momentum
relation for the photon
$$E^2 - p^2 c^2 = 0$$
and introduce matrices given in
$$S_x = \left(\begin{array}{ll}
0 \quad 0 \quad 0\\
0 \quad 0 \quad -\imath\\
0 \quad \imath \quad 0
\end{array}\right)\, , S_y = \left(\begin{array}{ll}
0 \quad 0 \quad \imath\\
0 \quad 0 \quad 0\\
-\imath \quad 0 \quad 0
\end{array}\right)\, ,$$
\begin{equation}
S_z = \left(\begin{array}{ll}
0 \quad -\imath \quad 0\\
\imath \quad 0 \quad 0\\
0 \quad 0 \quad 0
\end{array}\right)\, , I^{(3)} = \left(\begin{array}{ll}
1 \quad 0 \quad 0\\
0 \quad 1 \quad 0\\
0 \quad 0 \quad 1
\end{array}\right)\, ,\label{ed6}
\end{equation}
from which we get
$$\left(\frac{E^2}{c^2} - {\bf p^2}\right) {\bf \Psi} = \left(\frac{E}{c}I^{(3)} +
{\bf p \cdot S}\right){\bf \Psi}$$
\begin{equation}
- \left(\begin{array}{ll}
p_x\\
p_y\\
p_z
\end{array}\right) \, ({\bf p \cdot \Psi}) = 0,\label{ed7}
\end{equation}
where $\Psi$ is a three component wave function and in general bold letters denote vector
quantization.\\
Equation (\ref{ed7}) implies
\begin{equation}
\left(\frac{E}{c} I^{(3)} + {\bf p \cdot S}\right) {\bf \Psi} =
0,\label{ed8}
\end{equation}
\begin{equation}
{\bf p \cdot \Psi} = 0,\label{ed9}
\end{equation}
where $S$ is given in (\ref{ed6}). There is also an equation for
${\bf \Psi^*}$ namely
\begin{equation}
\left(\frac{E}{c}I^{(3)} - {\bf p \cdot S}\right) {\bf \Psi^*} =
0,\label{ed10}
\end{equation}
\begin{equation}
{\bf p \cdot \Psi^*} = 0,\label{ed11}
\end{equation}
It is then easy to verify (Cf.ref.\cite{gersten}) that with the substitution of the usual
Quantum Mechanical energy momentum operators, we recover equations (\ref{ed3}) to (\ref{ed5})
for ${\bf \Psi}$ and its complex conjugate.\\
Recently a similar analysis has lead to the same conclusion. In fact
it has been shown that under a Lorentz boost \cite{dvg},
\begin{equation}
\left(\begin{array}{ll}
\Psi'\\
\Psi^{*'}\end{array}\right) = \left(\begin{array}{ll}
1 - \frac{({\bf S \cdot p})}{mc} + \frac{({\bf S \cdot p})^2}{m(E+mc^2)} \quad \quad 0\\
0 \quad \quad \quad \quad 1 + \frac{({\bf S \cdot p})}{mc} +
\frac{({\bf S \cdot p})^2}{m(E+mc^2)}\end{array}\right)
\left(\begin{array}{ll}
\Psi\\
\Psi^* \end{array}\right)\label{ed12}
\end{equation}
We would like to point out that equations (\ref{ed4}), (\ref{ed5}),
(\ref{ed8}) to (\ref{ed12}) display the symmetry
$${\bf p} \to -{\bf p} \quad , \Psi \to \Psi^*$$
We now invoke the Weinberg-Tucker-Hammer formalism (Cf.\cite{dvg})
which gives, for a Lorentz boost equations
\begin{equation}
\phi'_R = \left\{ 1 + \frac{{\bf S \cdot p}}{m} + {({\bf S \cdot
p})^2}{m(E + m)}\right\} \phi_R,\label{ed13}
\end{equation}
\begin{equation}
\phi'_L = \left\{ 1 - \frac{{\bf S \cdot p}}{m} + {({\bf S \cdot
p})^2}{m(E + m)}\right\} \phi_L,\label{ed14}
\end{equation}
where the subscripts $R$ and $L$ refer to the states of opposite helicity, that is left
and right polarised light in our case.\\
We now observe that equations (\ref{ed12}) and
(\ref{ed13})-(\ref{ed14}) are identical, but there is a curious
feature in both of these, that is that the photon of
electromagnetism is now seen to have a mass $m$.\\
It can be argued that this is a result of the non commutativity of
spacetime at a micro scale. We observe that a photon mass would
imply the equation
\begin{equation}
\partial^{\mu} F_{\mu \nu} = - m^2 A_{\nu},\label{ed15}
\end{equation}
where we have the usual equations of electromagnetism
\begin{equation}
\partial^{\mu} A_{\mu} = 0,\label{ed16}
\end{equation}
\begin{equation}
F_{\mu \nu} = \partial_{\mu} A_{\nu} - \partial_{\nu}
A_{\mu}\label{ed17}
\end{equation}
We note that from (\ref{ed17}) we get
\begin{equation}
\partial^{\mu} F_{\mu \nu} = D A_{\nu} - \partial^{\mu} \partial_{\nu} A_{\mu}\label{ed18}
\end{equation}
where $D$ denotes the D'Alembertian. In view of (\ref{ed16}), the
second term on the right side of (\ref{ed18}) would vanish, provided
the derivatives commute. In this case we would return to the usual
Maxwell equations. However in the non commutative case this extra
term is
$$p^2 A_\mu \sim m^2 A_\mu$$
remembering that we are in units $c = 1 = \hbar$.\\
Thus because of the non commutativity we get (\ref{ed15}) instead, of the usual Maxwell
equation.\\
The question of non commutativity and mass generation in the context of gauge theory has
been studied by the author elsewhere (Cf.\cite{annales,ijmpe2,uof}).\\
In any case the above points to the fact that there would be no massless particles in
nature. The point is, that in an idealized situation in which the radius $R \to \infty$,
the mass $m_\gamma \to 0$.\\
\section{Discussion}
We had referred to the violation in Lorentz symmetry caused by the
mass of the photon. It may be mentioned that such violations can be
deduced from the author's fuzzy spacetime approach. Here we have a
modified dispersion relation $(c = 1 = \hbar)$
\begin{equation}
E^2 = m^2 + p^2 + \alpha l^2 p^4\label{ee1}
\end{equation}
where $\alpha$ is a dimensionless constant of order unity. (For
fermions, $\alpha$ is negative). This leads to a modification of the
Dirac and Klein-Gordon equations at ultra high energies
(Cf.ref.\cite{uof,ijtp,ultra}). With this, it has been shown by the
author that in the scattering of radiation, instead of the usual
Compton formula we have
\begin{equation}
k = \frac{mk_0 + \alpha \frac{l^2}{2}[Q^2 + 2mQ]^2}{[m + k_0 (1 -
cos \Theta )]}\label{ee2}
\end{equation}
where we use natural units $c = \hbar = 1, m$ is the mass of the
elementary particle causing the scattering, $\vec{k} , \vec{k}_0$
are the initial and final momentum vectors respectively and $Q = k_0
- k,$ and $\Theta$ is the angle between the incident and scattered
rays. Equation (\ref{ee2}) shows that $k = k_0 + \epsilon$, where
$\epsilon$ is a positive quantity less than or equal to $\sim l^2$,
$l$ being the fundamental length. It must be remembered that in
these units $k$ represents the frequency. The above can be written
in more conventional form as
\begin{equation}
h \nu = h\nu_0 [1 + 0(l^2)]\label{ee3}
\end{equation}
Equation (\ref{ee3}) effectively means that due to the Lorentz symmetry violation in
(\ref{ee1}), the frequency is increased or, the speed of propagation of a given frequency
is increased. As noted such models in a purely phenomenological context have been considered
by Glashow, Coleman, Carroll and others. In any case what this means in an observational
context is that higher frequency gamma rays should reach us earlier than lower frequency
ones in the same burst. As Pavlopoulos reports (Cf.ref.\cite{pav}) this indeed seems to be
the case.\\
Subject to further tests and confirmation, for example by NASA's
GLAST satellite to be launched shortly \cite{glast}, spacetime at a
micro or ultra high energy level is not a smooth manifold, brought
out by this manifestation in for example (\ref{ee1}).\\
It may also be mentioned that further encouragement for these
approaches comes from the fact that in the AGASA experiment, already
$12$ events of the violation of the $GZK$ cut off have been
reported.


\begin{thebibliography}{99}
\bibitem {bhtd} Sidharth, B.G., Found.Phys.Lett., 19 (1), 2006, pp.87ff.
\bibitem {vigier} Vigier, J.P., IEEE Transactions on Plasma Science,
Vol.18, No.1, 1990, 64-72.
\bibitem {bartlett} Bartlete, D.F., and Corle, T.R., {\it Phys.Rev.Lett,}, Vol.55, (1985), p.49.
\bibitem {lakes} Lakes, R., Phys.Rev.Lett., Volume 80, Number 9,
1998 1826-1829.
\bibitem {cms} Cardone, F., Mignani, R., and Scrimaglio, R.,
Foundation of Physics, Vol.36, No.2, 2006, 263-290.
\bibitem {pav} Pavlopoulos, T.G., {\it Phys.Lett.B}, 625, (2005), pp.13-18.
\bibitem {greiner2} Greiner, W., Muller, B., Rafelski, J., ``Quantum Electrodynamics of Strong Fields'', Springer-Verlag, Berlin, 1985.
\bibitem {bgs} Sidharth, B.G., PhD Thesis, Calcutta University, 1977.
\bibitem {mp} Sidharth, B.G., A Note on Massive Photons
\bibitem {ben} Bennett, C.L., et al., {\it Astrophys. J. Suppl}. {\bf 148}, (2003)
1.
\bibitem {hin} Hinshaw, G., et al., {\it Astrophys. J. Suppl}. {\bf 148}, (2003), 135.
\bibitem {mer} Mersini-Houghton, L., {\it Mod.Phys.Lett.A.}, Vol.21, No.1, (2006), 1-21.
\bibitem {fpl1} Sidharth, B.G., {\it Found.Phys.Letters.}, 15 (6), (2002), 577-583.
\bibitem {fpl2} Sidharth, B.G., {\it Found.Phys.Letters.}, 17 (5), (2004), 503-506.
\bibitem {ijmpe} Sidharth, B.G., {\it Int.J.Mod.Phys.E,} 15(1), (2006), pp.255ff.
\bibitem {uof}   Sidharth, B.G., {\it The Universe of Fluctuations} (Springer, NL, 2005).
\bibitem {ig} Ignatiev, A. Yu., and Joshi, G.C., Mod.Phys.Lett.A., 11, 1996, pp.2735-2741.
\bibitem {land} Landsberg, P.T., {\it  Am.J.Phys.}, {\bf 51}, (1983), 274-275.
\bibitem {found} Sidharth, B.G., {\it Found.Phys.Letts}, 17 (5), (2004), 503-506.
\bibitem {evans} Evans, M., and Vigier, J.P., {\it The Enigmatic Photon} (Kluwer Academic, Dordrecht, 1995),
p.136.
\bibitem {neeman} Newman, E.T., {\it J.Math.Phys.}, 14 (1), (1973), pp.102-103.
\bibitem {gersten} Gersten, A., {\it Found.Phys.Lett.}, 12 (3), (1999), pp.291-298.
\bibitem {dvg} Dvoeglazov, V.V., and Gonzalez, J.L.Q., Found.Phys.Lett., 19 (2), (2006), pp.195ff.
\bibitem {dvg} Dvoeglazov, V.V., and Gonzalez, J.L.Q.,
{\it Found.Phys.Lett.}, 19 (2), 2006, pp.195ff.
\bibitem {annales} Sidharth, B.G., {\it Annales de la Fondation Louis de Broglie}, 27 (2), (2002), pp.333ff.
\bibitem {ijmpe2} Sidharth, B.G., {\it Int.J.Mod.Phys.E.}, 14 (6), (2005), 923ff.
\bibitem {ijtp} Sidharth, B.G., {\it Int.J.Th.Ph.}, Vol.43 (9),
(2004), p.1ff.
\bibitem {ultra} Sidharth, B.G., {\it Int.J.Mod.Phys.E.}, 14 (6),
(2005), pp.927ff.
\bibitem {glast} http://glast.gsfc.nasa.gov/
\end{thebibliography}
\end{document}